\documentclass[preprint2]{aastex} 

\slugcomment{AJ Version 30/5}

\shorttitle{PARSEC I: First results}
\shortauthors{Andrei et al.}

\begin{document}


\title{Parallaxes of Southern Extremely Cool objects \\
     I: Targets, Proper motions and first results. }


\author{A.H. Andrei\altaffilmark{1,2,3}, 
          R.L.      Smart\altaffilmark{2},
          J.L.      Penna\altaffilmark{1},
          V. A.   d'Avila\altaffilmark{1,4},
          B.        Bucciarelli\altaffilmark{2},       
          J.I.B.    Camargo\altaffilmark{1},
          M. T.     Crosta\altaffilmark{2},
          M.        Dapr\`a\altaffilmark{2},
          B.        Goldman\altaffilmark{5},
          H.R.A.    Jones\altaffilmark{6},
          M.G.      Lattanzi\altaffilmark{2},
          L.        Nicastro\altaffilmark{7},
          D.N.      da Silva Neto\altaffilmark{8}        
     and  R.        Teixeira\altaffilmark{9}}

      \altaffiltext{1}{Observat\'orio Nacional/MCT, R. Gal. Jos\'e Cristino
        77, CEP20921-400, RJ, Brasil.}
        
      \altaffiltext{2}{ INAF/Osservat\'orio Astronomico di Torino, Strada
        Osservat\'orio 20, 10025 Pino Torinese, Italy}

      \altaffiltext{3}{ Observat\'orio do Valongo/UFRJ, Ladeira Pedro
        Ant\^onio 43, CEP20080-090, RJ, Brasil}
        
      \altaffiltext{4}{ UERJ, Faculdade de Oceanografia, R. S\~ao Francisco Xavier 524, CEP20550-900, RJ, Brasil}
        
      \altaffiltext{5}{Max Planck Institute for Astronomy, Koenigstuhl 17,
        D--69117 Heidelberg, Germany}

      \altaffiltext{6}{ Centre for Astrophysics Research, Science and
        Technology Research Institute, University of Hertfordshire, Hatfield
        AL10 9AB}
      
      \altaffiltext{7}{INAF/Istituto di Astrofisica Spaziale e Fisica Cosmica,
        Bologna, Via Gobetti 101, 40129 Bologna, Italy}

      \altaffiltext{8}{ Centro Universit\'ario Estadual da Zona Oeste,
        Av. Manuel Caldeira de Alvarenga 1203, CEP23070-200, RJ, Brasil}

      \altaffiltext{9}{Instituto de Astronomia, Geofísica e Ciências
        Atmosféricas, Universidade de São Paulo, Rua do Matão, 1226 - Cidade
        Universitária, 05508-900 São Paulo - SP, Brazil}


\begin{abstract}
  We present results from the PARallaxes of Southern Extremely Cool objects
  (PARSEC) program, an observational program begun in April 2007 to determine
  parallaxes for 122 L and 28 T southern hemisphere dwarfs using the Wide
  Field Imager on the ESO 2.2m
  telescope. The results presented here include parallaxes
  of 10 targets from observations over 18 months and a first version proper
  motion catalog. 

  The proper motions were obtained by combining PARSEC observations
  astrometrically reduced with respect to the Second US Naval Observatory CCD
  Astrograph Catalog, and the Two Micron All Sky Survey Point Source
  Catalogue.  The resulting median proper motion precision is 5mas/yr for
  195,700 sources.  The 140 0.3deg$^{2}$ fields sample the southern hemisphere
  in an unbiased fashion with the exception of the galactic plane due to the
  small number of targets in that region.  The proper motion distributions
  are shown to be statistically well behaved.  External comparisons are also
  fully consistent.  We will continue to update this catalog until the end of
  the program and we plan to improve it including also observations from the
  GSC2.3 database.

  We present preliminary parallaxes with a 4.2 mas median precision for 10
  brown dwarfs, 2 of which are within 10pc.  These increase by 20\%
  the present number of L dwarfs with published parallaxes.  Of the 10 targets, 7
  have been previously discussed in the literature: two were thought to be
  binary but the PARSEC observations show them to be single, one has been
  confirmed as a binary companion and another has been found to be part of a
  binary system, both of which will make good benchmark systems. These results
  confirm that the foreseen precision of PARSEC can be achieved and that the
  large field of view will allow us to identify wide binary systems.

  Observations for the PARSEC program will end in early 2011 providing 
  3-4 years of coverage for all targets. The main expected outputs are:
  more than a 100\% increase of the number of L dwarfs with parallaxes; to
  increment - in conjuction with published results - to at least 10 the number
  of objects per spectral subclass up to L9, and; to put sensible limits on
  the general binary fraction of brown dwarfs. We aim to contribute
  significantly to the understanding of the faint end of the H-R diagram and
  of the L/T transition region.
\end{abstract}





\section{Introduction}

The first brown dwarf, GD 165B, was discovered in 1988
(\citeauthor{1988Natur.336..656B}) but was not recognised as such until 1995
when Gl229B (\citeauthor{1995Natur.378..463N}) and other objects with the same
characteristics were found. Rapidly many examples were discovered primarily in
the large infrared surveys, i.e. the Two Micron All Sky Survey
\citep[2MASS,][]{2006AJ....131.1163S} and Deep Near-Infrared Survey
\citep[DENIS,][]{1999AA...349..236E}, and, the deep optical Sloan Digital Sky
Survey \citep[SDSS,][]{2000AJ....120.1579Y}. It was soon realized that new
spectral types, L and T, were needed \citep{1999ApJ...519..802K}. Since then
over 700 L/T dwarfs have been discovered (www.dwarfarchives.org 2/2010) by
various authors and just recently a sample of 210 new L dwarfs from the
SDSS was announced  \citep{2010AJ....139.1045S}. These objects have heralded
a whole new sub-field of astronomy. 

Interest in brown dwarfs has been particularly prominent in the interpretation
of their spectral and photometric properties. Theory has been led in
unexpected directions by unpredicted behaviors: the very strong evolution of
spectral type with age \citep{1997ApJ...491..856B}; the ``hump'' in the J band
magnitude as a function of spectral type at the L/T transition
\citep{2003AJ....126..975T}; notable differences between infrared spectra of
optically classified objects and vice-versa \citep[e.g. fig 3
in][]{2008ASPC..384...85K}; a turnaround in color at the T8/T9 spectral type
\citep{2007MNRAS.381.1400W}. We are also slowly uncovering significant numbers
of L and T sub-dwarfs \citep{2009ApJ...694L.140S, 2004ApJ...614L..73B,
  2009ApJ...706.1114B, 2010ApJ...710...45B} and other peculiar objects that
challenge the theoretical models.

Parallax is a crucial parameter for understanding these objects as it is the
only direct way to calculate an absolute magnitude and hence energetic
output. In brown dwarf structure models, particularly for T dwarfs, the
determination of metallicity and surface gravity from spectra is degenerate
\citep{2009arXiv0901.4093L} and hence luminosity, which requires a parallax,
is used to constrain either the radius or the temperature and help break this
degeneracy. Precise absolute velocities, that in turn provide age and origin
indications, require precise parallaxes.   

\begin{figure}
\plotone{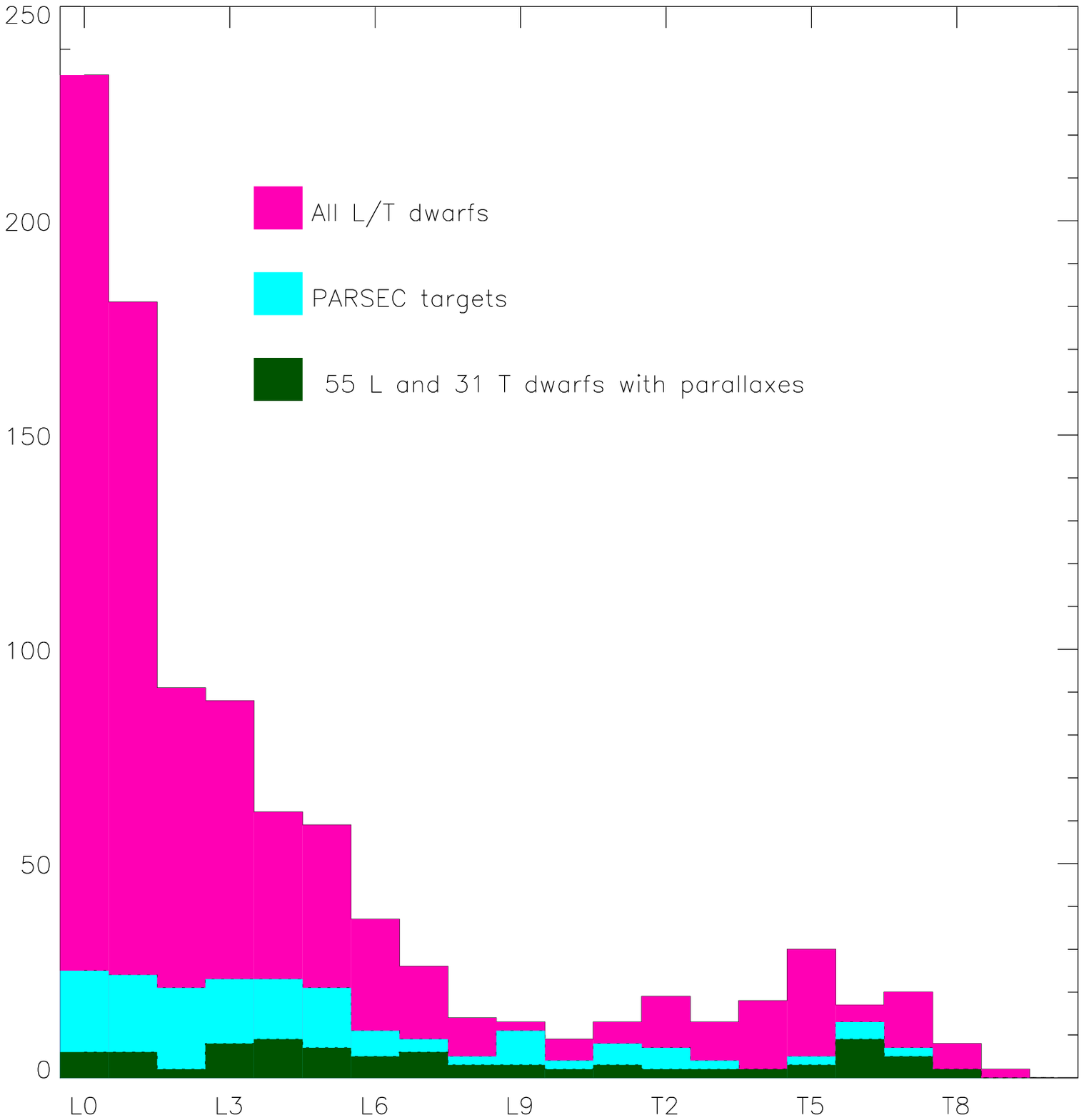}
\caption{The distribution of the 752 known L and T dwarfs as of 2/2010
    (www.dwarfarchives.org). Overplotted the distribution of the 90 objects with
    published parallaxes and the 140 objects in the PARSEC program as indicated in
    the legend.}
  \label{looklist.cps_page_8}%
\end{figure}

In light of the role of distance it is important that for these new objects we
have a significant sample with measured parallaxes. As shown in Figure
\ref{looklist.cps_page_8}, only a small fraction of known L/T dwarfs have
measured parallaxes - the black histogram - which limits any calibrations and
generalizations we can make. To increase the current sample the Osservatorio
Astronomico of Turin and Observat\'orio Nacional of Brazil begin in 2007 the
PARarallaxes of Southern Extremely Cool objects (hereafter PARSEC) program to
determine parallaxes for 140 bright L and T dwarfs. In Figure
\ref{looklist.cps_page_8} we include the PARSEC objects that illustrate our
goal to attain at least 10 objects per spectral bin for L dwarfs and to
increase the current sample for the fainter T dwarfs. A number of other
programs are also underway to address this shortfall (for example C. G. Tinney
private communication 
and J. K. Faherty as part of the Brown Dwarf Kinematics Project 
 \citep{2009AJ....137....1F}) but even including the expected additional 
objects the  results of the PARSEC program will at least double number of 
L and bright T dwarfs with parallaxes.


In this paper we present the PARSEC program: section 2 describes the
instrument and target selection; section 3 details the observational and
reduction procedures, and section 4 the procedures used to produce a catalog
of standard proper motions and preliminary parallax solutions for 10
objects. Finally in section 5 we discuss some uses we have made of this
catalog and future plans.

\section{The PARSEC observational program}
\subsection{The instrument}
The primary instrument for this program is the Wide Field Imager
\citep[WFI,][]{1999Msngr..95...15B} on the ESO 2.2m telescope. This is a mosaic
of 8 EEV CCD44 chips with 2k$\times$4k 15 $\mu$m pixels, providing a total field
of 32.5 by 32.5 arcmin. This instrument/telescope combination
was chosen for a number of reasons:
\begin{enumerate}
\item The instrument is fixed and stable, both crucial requirements in relative
  astrometry work
\item The plate scale of 0.2''/pixel is optimal for this work
as it offers better than Nyquist sampling even in the best seeing.
\item The field size of 0.3 square degrees allows a reasonably thorough
  search for nearby companions
\item It already has a proven track record for the determination of parallaxes
  of dwarf objects \citep{2007AA...470..387D}
\end{enumerate}
It was decided to observe in the $z$ band (Z+$/$61 ESO\#846, central
wavelength 964.8nm, FWHM 61.6nm) which was a compromise between the optimal QE
of the system in the $I$ band, and the expected brightness of the targets
which have a $I-z$ of about 2. To keep the exposure times within 300s we
observed only objects brighter than $z<$19.

\subsection{Observations}
The observational procedure is as follows. 
\begin{enumerate}
\item For each field we make one quick 50s exposure and locate the target.
\item Using the WFI move-to-pixel procedure we offset the telescope to move
  the target to pixel (3400,3500), which is in a flat part of CCD\#7
  (Priscilla) at less than 1/4 of the diagonal leading to the optical center
  of the mosaic.
\item We make the first science exposure of 150s for objects with $z<$18.0 and 300s for $z\ge$18.0.
\item The camera is then slightly offset, 24 pixels in both directions, and the
  second science image of the same exposure time is automatically begun.
\item We check the counts of the target in the first image. If the
  signal-to-noise of the target in the first image is less than 100, in real
  time we increased the exposure time accordingly. This is usually only the
  case in particularly poor sky conditions.
\end{enumerate}
This procedure is very efficient and the dead-time for the telescope is
minimal.  The total time for a target is 10-25 minutes depending on 
magnitude and other overheads, enabling us to observe 3-4 objects per
hour. Our time allocation usually results in always having grouped nights
and, as multiple observations in the same run are of limited value, to both
increase the sample and allow some redundancy, the target list has 6-8 objects
per hour. We attempted to observe the majority of targets close to the meridian
except during the twilight hours when we wish to include rising or setting
targets at their maximum parallactic factor.

Observations began in April 2007 and, as of September 2009, targets have between
1.5 and 2.5 years of observations.  The frequency of observations have been
reasonably constant with a 2-3 night run every two months. 

\begin{figure}
\plotone{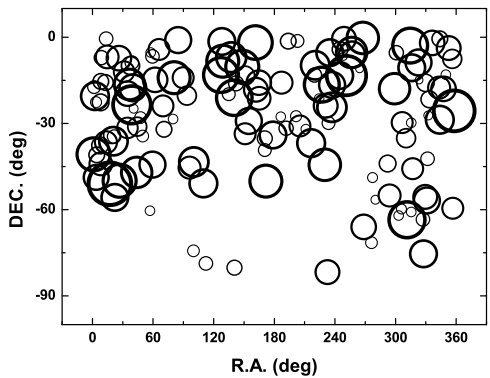}
\caption{The equatorial coordinates distribution of the 140 sources of program. 
The size of the circles is in proportion to the object $z$ magnitude. 
Notice that all targets belong in the austral hemisphere, 
and that they are clear of the galactic disk. 
    }
  \label{SkyDistribution}%
\end{figure}

Figure \ref{SkyDistribution} displays the sky distribution of the targets. 
Table \ref{DATES} summarizes the dates and nights of the observations taken up 2009.

\begin{table}[ht]
  \caption{Observations of the PARSEC program at the ESO2p2/WFI up to 2009.}
\begin{tabular}{ll}
\hline  \\
Date & Nights \\
\tableline
 April 2007   & 09, 10, 11, 12\\
 August 2007   & 31\\
 September 2007   & 01, 02\\
 October 2007   & 05, 06, 07\\
 January 2008   & 04, 05\\
 February 2008   & 26, 27\\
 April 2008   & 02, 03\\
 May 2008   & 27, 28\\
 August 2008   & 21, 22\\
 October 2008   & 24, 26\\
 December 2008   & 18, 20\\
 March 2009   & 01, 02, 03\\
 April 2009   & 30\\
 May 2009   & 06, 08\\
 July 2009   & 21, 22, 23\\
 December 2009   & 15, 18, 21\\
 \hline
\label{DATES}
\end{tabular}
\end{table}

\subsection{Target Selection}
The targets were selected using the following criteria:
 \begin{enumerate}
\item All southern L and T dwarfs discovered before April 2007 
\item Brighter than $z$=19
\item No more than 8 objects in any RA hour 
\item The brightest examples within each spectral bin
\item A uniform spectral class distribution
\item A photometric distance smaller than 50pcs.
\end{enumerate}
The photometric distances were estimated using the 2MASS magnitudes
transformed to the MKO system using \cite{2004PASP..116....9S}
and the color -
absolute magnitude compilation given in \cite{2004AJ....127.3553K}.
Exceptions were made to include any objects that were underrepresented, e.g.
most known T dwarfs are too faint for this program so any T dwarf with z $<$
19 was given high priority. By applying the above criteria and by removing
those objects we were not able to observe during the first runs due to time
compression, the remaining list has 140 targets as shown in table
\ref{targets}. Listed are: the 2MASS counterpart name, shortened name used in
this paper, published $z$-band magnitude - if no published value is available
this is estimated from the J band magnitude and spectral type, 2MASS
magnitudes, nominal spectral type and the discovery name. Most of the objects
were chosen from the dwarfarchives.org while some are from the catalogs of
\citet{2007AA...468..163D} and \citet{2004AA...421..763P}.

\addtocounter{table}{1}

\subsection{Image Reduction Procedures}

The bias, dark and flat image corrections followed standard procedures while
fringe removal required a tailored approach. The interference fringes in the
WFI $z$-band image are severe, an examination of the counts shows they can
vary by up to 10\% over the distance of a few pixels. Fringing is an additive
effect that can be corrected making a fringe map and subtracting it from the
raw images. The suggested WFI approach is to apply a standard fringe map which
is updated at periodic intervals. We found it improved our centroiding by
adopting a different approach and to understand why we first consider the
cause of fringing.

Fringes are caused by the constructive and destructive interference of the
night sky emission lines that are reflected from the bottom of the CCD silicon
layer with incoming radiation. Fringes are time and observation dependent for
a number of reasons e.g.: changes in the brightness of the night sky emission
lines, changes in the thickness of the silicon layer which is a function of
the temperature of the CCD, changes in the angle of incidence of the light on
the CCD which is a function of flexure. The ideal case would therefore be to
make a fringe map for each image but this is not feasible. Our compromise is
to make a nightly fringe map whenever possible.

The general procedure to construct a fringe map is to mask out objects then 
build a mean map from all of the observations in a given night scaled
appropriately to reveal the fringe signal. Specifically we followed the
following steps:
\begin{enumerate}
  \item For all images we identify all the objects and make an object mask.
  \item For each image we make a sky map by fitting a plane to all the
    unmasked pixels including a 3$\sigma$ clipping rejection criteria. This
    changes in the course of the night so it is necessary to remove it from
    each frame independently.
  \item We select a fringe calibration image subset consisting of all the
    short 50s and 4 of the long science exposures. We did not include all the
    science images in this subset as the object mask does not always cleanly
    block out all of the target signal and using all the science frames with
    the target on the same pixel results in a ghost image around the
    move-to-pixel position.
  \item We make a median image by scaling all subset images by the exposure
    time and making a median of the unmasked pixels.
  \item The first fringe map is constructed by smoothing the median image
    using a block size of 5 pixels.
  \item This first fringe is subtracted from all images providing sky
    subtracted and relatively fringe free observations.
  \item We make a new median image scaling the cleaned subset images by the
    weighted mean difference between  the input image and the fringe image.
  \item We construct a new fringe map smoothing the median image and then
    apply it to all the cleaned images providing fringe-free images.
\end{enumerate}

In the first iteration we use
the exposure time as a scale factor as the fringing will systematically affect
the mean image counts, in the second the majority of the fringes are removed
and we use the mean count as the scale factor which reflects the overall sky
conditions as well. Below we discuss the effect of this fringing on the
centroiding.

\subsection{Centroiding and Feasibility Tests}
\begin{figure}
\plotone{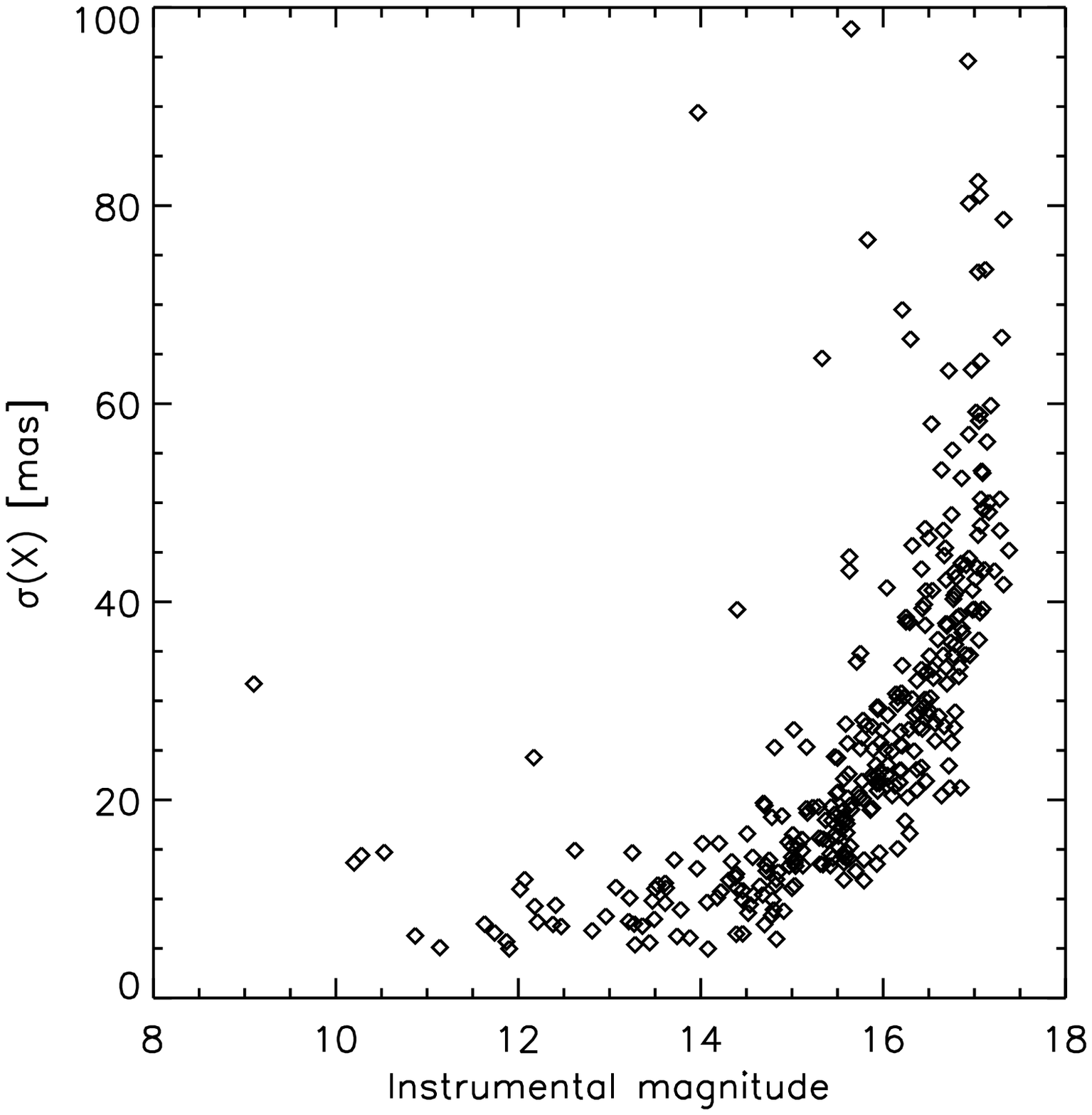}
\caption{The root-mean-square of the X coordinates residuals for stars in common with
    a 0719-50 image sequence spanning 1.5 years as a
    function of $z$ magnitude. The centroids were all derived using the TOPP 
    two dimensional gaussian fit.} 
  \label{f5700007rr3_fortest4b.ps_page_1}%
\end{figure}

The WFI, having a large field of view, has significant astrometric distortions
and the CCDs have significant relative tilts (see the WFI section at
www.eso.org).  However, the fundamental requirements for relative astrometry,
that underlie {\bf{all}} small field parallax determinations, are stability and
repeatability.  For this reason we use the WFI move-to-pixel routine to put 
the target on the same pixel for each
science exposure and only consider astrometric distortion changes over the
observational campaign. 
The move-to-pixel position, (3400,3500), is sufficiently inside CCD\#7 that
reference stars from only the top third of the chip are needed to make a
low-noise astrometric transformation between different epochs. The
move-to-pixel procedure introduces a significant overhead but as shown by
\citet{2002AJ....124..601P}, on a similar mosaic, the chips move relative to
one another and this introduces a change in the astrometric
deformation that was impossible to model at the mas level as required by
our parallax goal.

We have tested three centroiding routines: a two dimensional Gaussian fit to
the psf as used in the Torino Observatory Parallax Program
\citep[TOPP]{SMA99A}, a one-dimensional Gaussian fit to the marginal
distributions, and the Gaussian psf fit provided by daophot in IRAF. In a
comparison of object positions observed on 14 nights over an 18-month period
of the field around the object 0719-50 we found the TOPP
routine worked best. In figure \ref{f5700007rr3_fortest4b.ps_page_1} we plot
rms of the position differences as a function of instrumental $\it{z}$ magnitude for the frames
in question. The median rms for all objects is 23mas and for the brighter
objects from 12-16 is 10mas.  If we apply the same test on images that have
not been fringe-corrected as described above we find the median precision has
deteriorated to 28mas.  Applying the fringe map supplied by the WFI
calibration team provided a precision that was intermediate between a nightly
fringe correction and no correction.

\section{Parallaxes}

\begin{figure}
\plotone{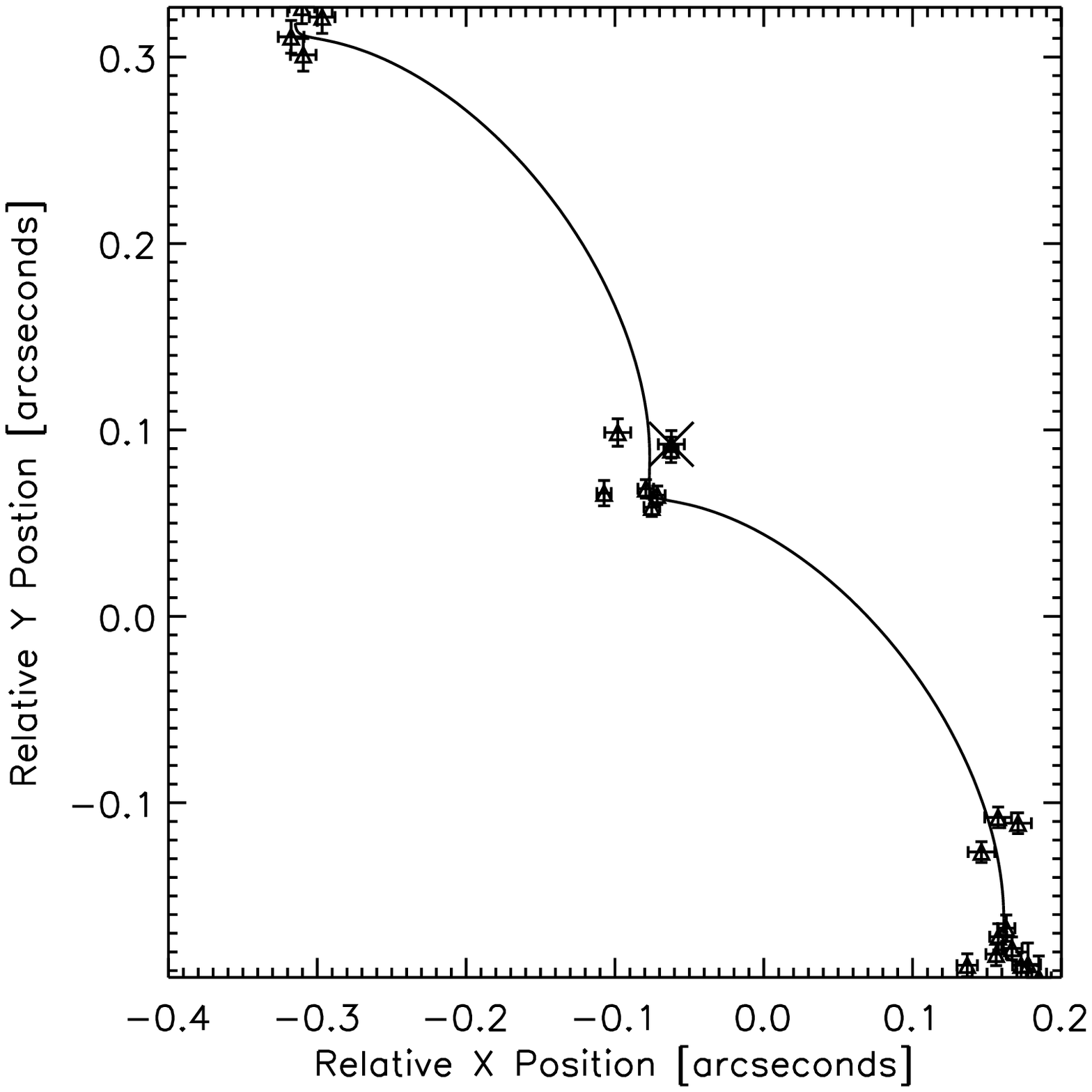}
\plotone{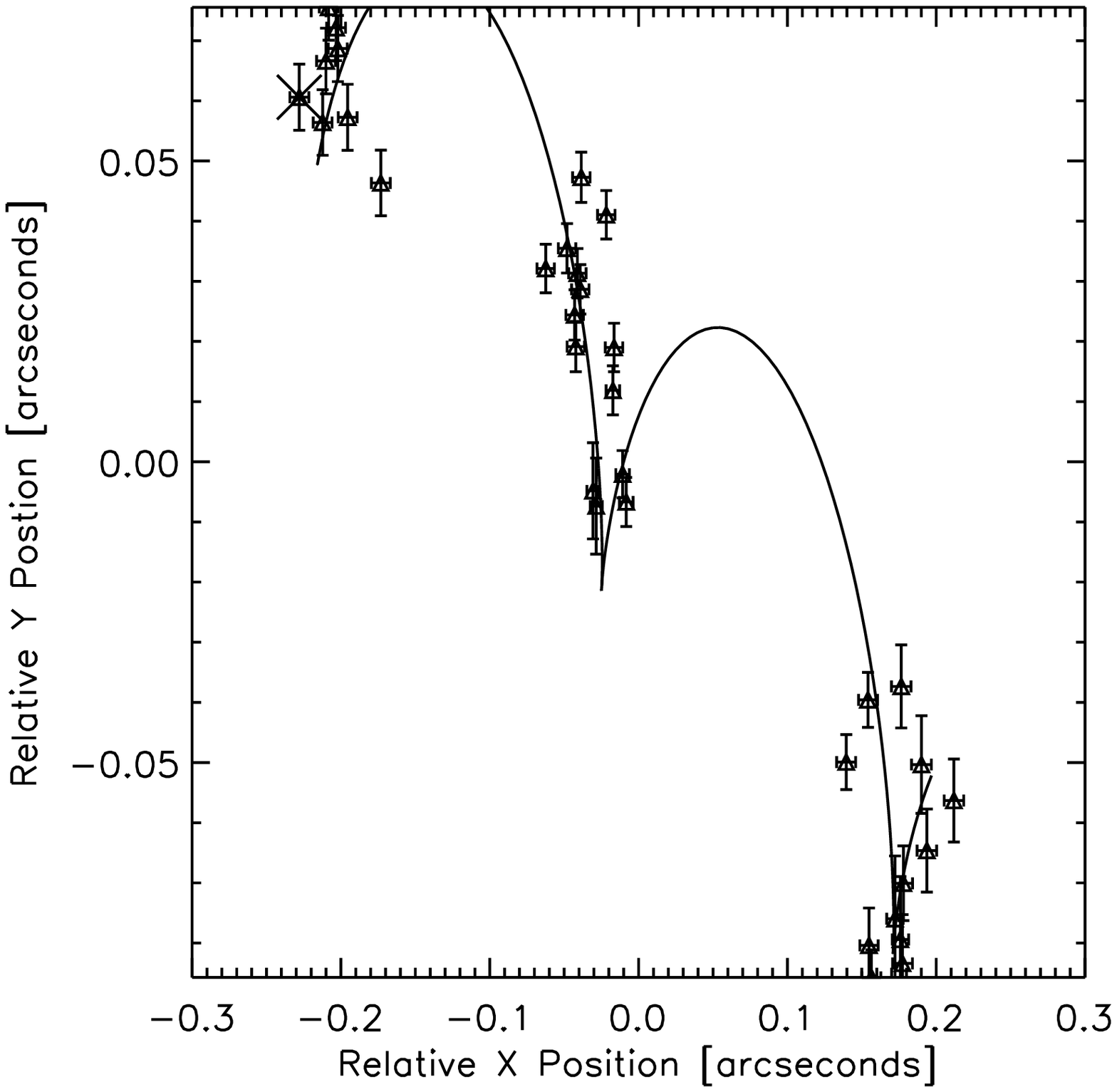}
\caption{Observations for two example targets 0719-50 and 1004-33.
}
  \label{twosolutions}%
\end{figure}

\begin{table*}[ht]
  \caption{Parallaxes and proper motions for a sample of PARSEC L-dwarfs.}
\begin{tabular}{rrrrrrrrr}
\hline  \\
ID &  $\alpha$ ~~~     & $\delta$ ~~~     &$N_*,N_f$ &  $\pi$          & $\mu_{\alpha}$      & $\mu_{\delta}$      & $\Delta$T &  COR \\
   &  h:m:s ~~   & d:':'' ~~   &          &     mas         &   mas/yr          &   mas/yr           & yrs      &  mas   \\
\hline  \\                                                                                                                     \\
0539-00 &  5:39:51.9 & - 0:58:58.3 &  31, 12 &    82.0 $\pm$  3.1 &   157.0 $\pm$  4.8 &   321.6 $\pm$  3.9 &   1.40 &   1.13 \\
0641-43 &  6:41:18.5 & -43:22:28.0 &  14, 29 &    55.7 $\pm$  5.7 &   215.9 $\pm$  8.9 &   612.8 $\pm$  9.0 &   1.95 &   1.00 \\
0719-50 &  7:19:32.0 & -50:51:41.3 &  22, 34 &    32.6 $\pm$  2.4 &   198.1 $\pm$  3.2 &   -61.4 $\pm$  3.9 &   1.98 &   0.90 \\
0835-08 &  8:35:42.2 & - 8:19:21.7 &   9, 20 &   117.3 $\pm$ 11.2 &  -519.8 $\pm$  7.7 &   285.4 $\pm$ 10.5 &   1.96 &   1.08 \\
0909-06 &  9:09:57.3 & - 6:58:18.8 &  20, 23 &    42.5 $\pm$  4.2 &  -184.0 $\pm$  2.5 &    20.7 $\pm$  3.0 &   2.08 &   1.19 \\
1004-33 & 10:04:39.5 & -33:35:21.9 &  16, 22 &    54.8 $\pm$  5.6 &   243.5 $\pm$  4.0 &  -253.2 $\pm$  3.4 &   2.06 &   9.51 \\
1018-29 & 10:18:58.5 & -29:09:54.2 &  32, 23 &    35.3 $\pm$  3.2 &  -340.8 $\pm$  1.8 &   -94.0 $\pm$  2.7 &   2.08 &   1.01 \\
1539-05 & 15:39:42.1 & - 5:20:41.5 &  17, 18 &    64.5 $\pm$  3.4 &   603.1 $\pm$  2.6 &   105.0 $\pm$  3.4 &   2.06 &   1.12 \\
1705-05 & 17:05:48.4 & - 5:16:46.9 &  96, 17 &    44.5 $\pm$ 12.0 &   110.9 $\pm$ 12.1 &  -115.5 $\pm$  7.1 &   1.98 &   0.59 \\
1750-00 & 17:50:24.5 & - 0:16:13.6 &  29, 39 &   108.5 $\pm$  2.6 &  -398.3 $\pm$  3.1 &   195.3 $\pm$  3.4 &   2.08 &   0.56 \\
\label{parallax}
\end{tabular}
\tablecomments{$N_*$ = number of reference stars, $N_f$ = number of frames, $\Delta$T = epoch
range, COR = correction to absolute parallax.}
\end{table*}

To determine parallaxes for our targets we consider only the top third of
CCD\#7. This is sufficiently large that, at these magnitudes, we have enough
reference objects for a transformation to a common system and sufficiently
small we can assume the variation in astrometric distortion over the
observational campaign is smaller than the errors of a linear
transformation. Once the (x,y) coordinates have been determined, the parallax
and proper motions are derived using the methods adopted in TOPP
\citep{Sma03a, 2007AA...464..787S}.  Schematically, we transfer the base frame
to a standard coordinate system using the UCAC2 stars, adjust all subsequent
frames to this base frame using all common stars and a simple linear
transformation, and find the relative parallax and proper motion of the target
star by a fit to the resulting observations in the system of the base
frame. In the $z$-band the atmospheric refraction is small and we assume the
differential color refraction to be negligible \citep{2002PASP..114.1070S}. To
calculate the correction from relative to absolute parallax we use the galaxy
model of Mendez \& van Altena (1996\nocite{Mendez1996}) in the $z$ band to
estimate the mean distance of the common stars. For the reference stars in
these fields this parallactic mean distance is of the order of 1mas with an
error of $<$30\% . For more details on the parallax determination procedure
and this correction the reader is refered to \cite{Sma03a,
  2007AA...464..787S}.

In figure \ref{twosolutions} we reproduce the solutions for two 
targets, 0719-50 and 1004-33. These two objects were also found to have
companions as discussed below. In Table \ref{parallax} we report the
parallaxes for 10 objects observed in the early runs of the PARSEC
program. Listed are: object ID, position, number of stars, number of frames, absolute
parallax, absolute proper motions, epoch difference and correction applied
from relative to absolute parallax. In the following we discuss some of these objects in more detail.

{\it \object{0539-00} }
This object was found to have a photometric variability on a timescale of
13.3h \citep{2001AA...367..218B}, we note in our sequence the instrumental
magnitude decreases by 0.05$\pm$0.03 magnitudes over the two year
sequence. The radial velocity was found not to vary for two observations
spaced 4 years, excluding a companion of mass greater than 10 M$_J$
\citep{2007ApJ...666.1205Z}; the astrometric residuals also show no evidence
for binarity.

{\it \object{0719-50} } For this object we find
$\mu_{\alpha},\mu_{\delta}$ = 198.1 $\pm$ 3.2, -61.4 $\pm$ 3.9 mas/yr and it
is identified by the PARSEC observations as a common proper motion companion
of 2MASS07193535-5050523 with $\mu_{\alpha},\mu_{\delta}$ = (200.3 $\pm$ 8.9,
-67.6 $\pm$ 5.7). The parallaxes both agree within the errors confirming the
binary nature of this system. Both these objects have previous proper motion
estimates, 0719-05 of 199.11 $\pm$ 20.49, -46.440 $\pm$13.78 mas/yr
\citep{2008MNRAS.390.1517C} and 2MASS07193535-5050523 of 206.2,-64.2
\citep{2007AJ....133.2898F}, but they were not noted as a common proper motion
system. 
The analysis of the proper motion distributions in the range of magnitude and 
in the sky loci surveyed by the entire PARSEC program, gives a probability 
smaller than 0.002 for a chance occurance of such common pair of large proper 
motions. This chance becomes even smaller if the common distance is also 
considered.
The brighter star has GSC2.3 magnitudes of $B_J=$16.01, $R_f=$13.50
and $I_n$=11.60 and 2MASS magnitudes of $J$=10.33, $H$=9.74, and $K$= 9.482.
Combining the magnitudes and distance  with the calibrations in
\cite{2002AJ....123.3409H} we find that the most consistent spectral 
type for this object is a M3-M4 dwarf.

{\it\object{0835-08} }
\citet{2003AJ....126.2421C} find a spectroscopic distance of 
8.3pc in agreement with our distance of 8.5pc, the nearest target in this 
sample and one of nearest known  L dwarfs to date.  The astrometric residuals
present no evidence of binarity, which, combined with the consistency of the
photometric and parallactic distances, allows us to confidently say that this
is a single system. 

{\it\object{0909-06} }
This is considered the prototypical L0 object, marking the begining of the L
dwarf sequence with a temperature of 2200K \citep{2000ApJ...538..363B}.  We
expect by the end of the program to have the distance to this object to better
than 5\%.

{\it\object{1004-33} } We confirm, as suggested in \citet{2008MNRAS.390.1517C}
and \citet{2005AN....326..974S}, that this object is a binary companion of the
nearby bright object LHS5166. The proper motions and parallaxes of the two
objects are both within one sigma of each other.  The brighter star has GSC2.3
magnitudes of $B_J=$15.51, $R_f=$13.33 and $I_n$=11.29 and 2MASS magnitudes of
$J$=9.85, $H$=9.30, and $K$=9.03.  Combining the magnitudes and distance with
the calibrations in \cite{2002AJ....123.3409H} we find the spectral type for
this object is a M3 dwarf consistent with the dM4.5e found in
\citet{2005AN....326..974S}.

{\it {\object{1705-05} }}
In \citet{2006AJ....132..891R} they consider the possibility that this object
is part of a binary system with a companion at position angle of 5$^\circ$ and
distance 1''.36. This proposed companion was too faint to be frequently observed as
part of our program, however, the color indicates a spectral type of T1-T2
that is inconsistent with the spectral type indicated by the apparent
magnitude and distance of T7-T9. Hence, we conclude this object is more likely
to be a background late M dwarf at $\sim$200 parsecs rather than a companion to 1705-05.

{\it {\object{1750-00} }} 
Based on spectroscopic observations, \citet{2007MNRAS.374..445K} found a
distances of 8$^{+0.9}_{-0.8}$pc and due to a discrepancies in the spectral type
indicators suggest that it may be a binary system of L5-L6 and
L8-L9 dwarfs. We find the trigonometric distance is consistent with the
photometric one and do not find any evidence of binarity in the residuals
which implies it is a single system and the discrepancies must have some other
explanation.

{\it {\object{0641-43}, \object{1018-29} $\&$ \object{1539-05}}}
The first two objects are type L1, and the third is L2. They are all in the
20-30pc distance range. {\object{0641-43}} and {\object{1539-05}} are fast
moving objects, with one of the proper motion components larger than
600mas/yr. None of them was the subject of any particular discussion in the
literature.

\begin{figure}
\plotone{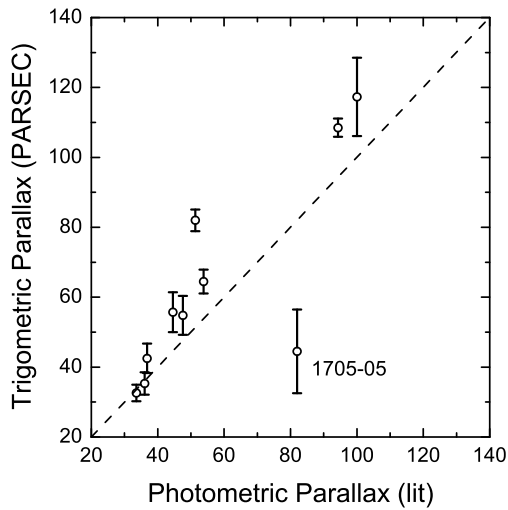}
\caption{Comparison between the preliminary parallaxes derived by the PARSEC program for the 10 targets in Table \ref{parallax} against the corresponding spectroscopic parallaxes. Notice how the photometric parallax of 1705-05 is affected by the superimposed nearby star (see text).
}
  \label{CompParallaxTrigXLit}%
\end{figure}

Figure \ref{CompParallaxTrigXLit} compares the trigonometric parallaxes from
Table \ref{parallax} against the corresponding spectroscopic parallaxes based
on the calibration of \citet{2004AJ....127.3553K}. The small number of sources
precludes any direct conclusion, yet the targets asymmetry relatively to the
equal values diagonal brings support to the importance of trigonometric
parallaxes as the fundamental calibrators for spectroscopic distances.

\section{Proper Motions}
The parallax determination of the targets uses only the upper third of CCD7;
however, the reduction pipeline is applied to the entire mosaic of 8
CCDs. From this data we have constructed a proper motion catalogue, sampling
the whole of the southern hemisphere with the exception of the lowest galactic
latitudes where the number of known L/T dwarfs is significantly reduced.  This
proper motion survey can be used to search for companions to the targets, and
other fast moving objects which will usually be nearby and/or sub-dwarfs.
Combined with the magnitudes, the proper motion survey can also be used to
build a reduced proper motion diagram to search for brown dwarf
candidates. This catalog contains proper motion determinations for 195,700
objects.

Independently for each CCD and each observation we have determined an astrometric
reduction relative to the Second US Naval Observatory CCD Astrograph Catalog
\cite[UCAC2,][]{2004AJ....127.3043Z}.  The average number of reference stars
was 20, with which polynomial functions were adjusted on right ascension and
declination. Depending on the number of reference stars the polynomial degree
was 2 or 3 and cross terms have been included. The rms errors of the solutions
did not show any dependence on the type of polynomial employed.

The proper motion determination was made from a match to the 2MASS point source
catalogue. In principle, the program frames should be complete with respect to
2MASS and the epoch difference is small, so a nearest neighbor match should
be sufficient to not mismatch high proper motion objects. As a safety measure,
the proper motions were determined for each observation combination and later
averaged. Deviant values were removed from these averages, as they either
came from unrecognised frame defects or faulty measurements. A more robust
algorithm is being developed using the GSC2.3 positions at different epochs.
At the targets galactic latitudes, blending is rare and its effects would be
negligible to the relatively bright 2MASS stars. 

\begin{figure}
\plotone{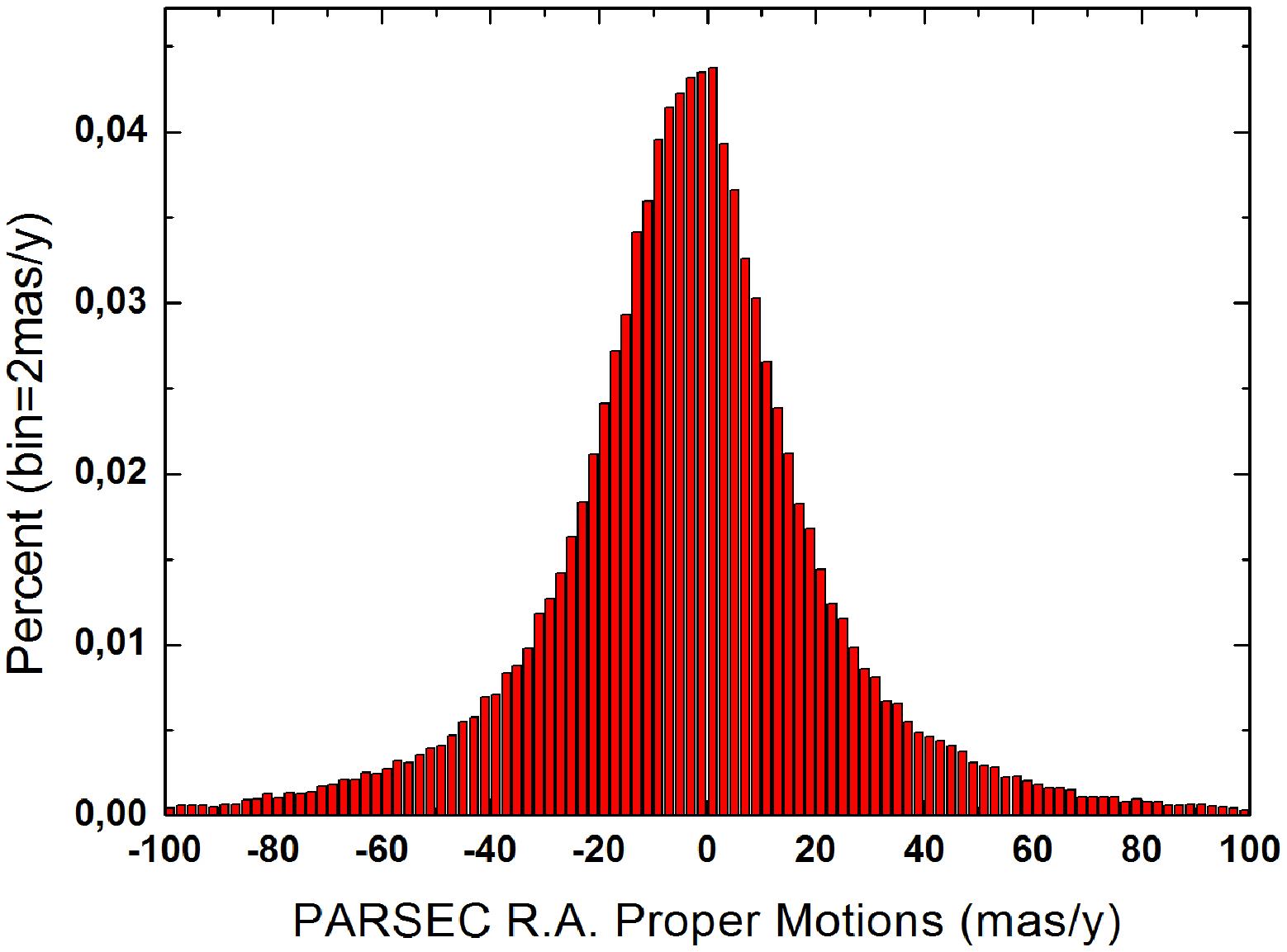}
\plotone{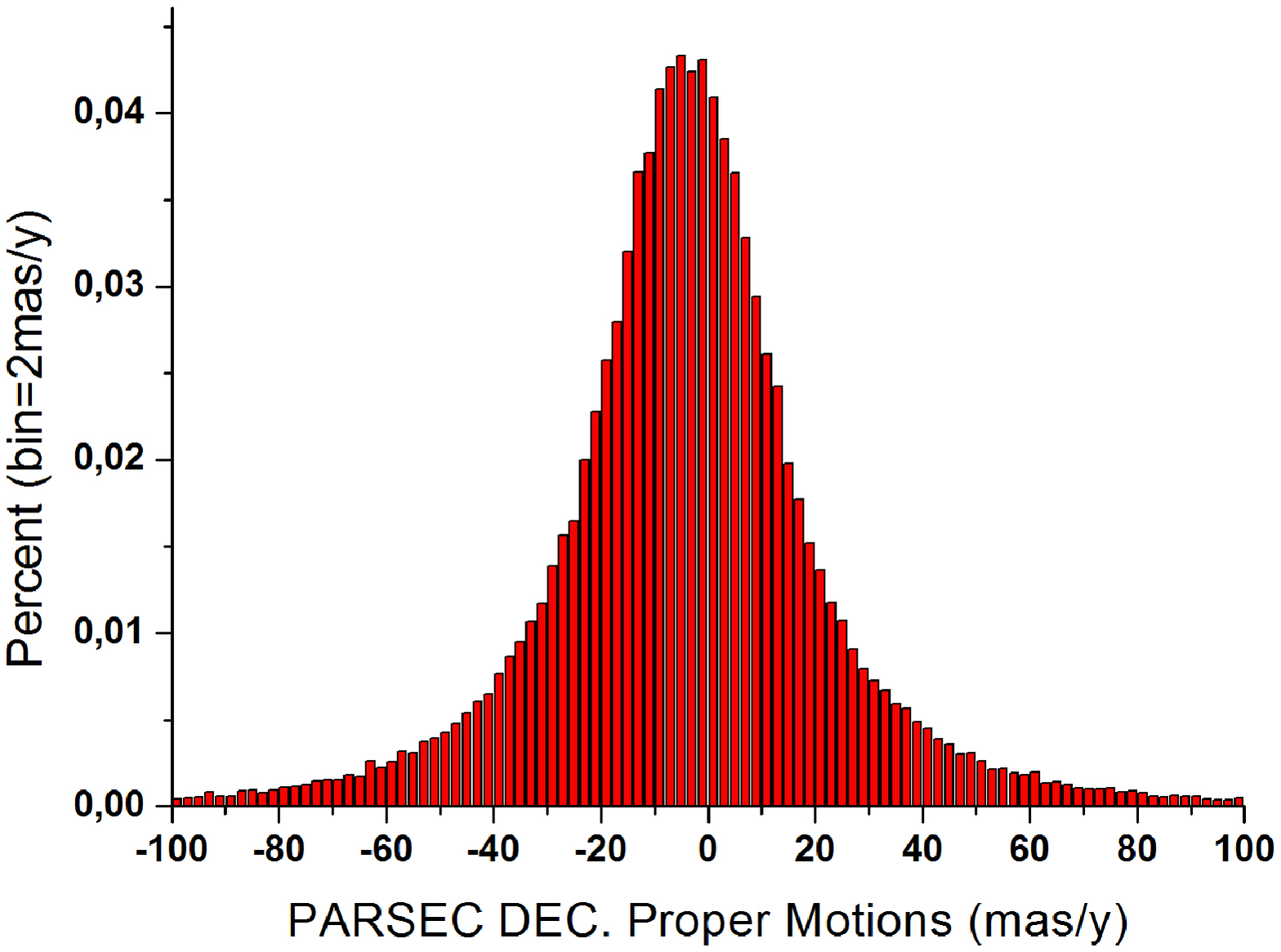}
\caption{ Proper motion distributions in right ascension and declination. }
  \label{Fig9a_PMRAdistribution.ps}
\end{figure}
\begin{figure}
\plotone{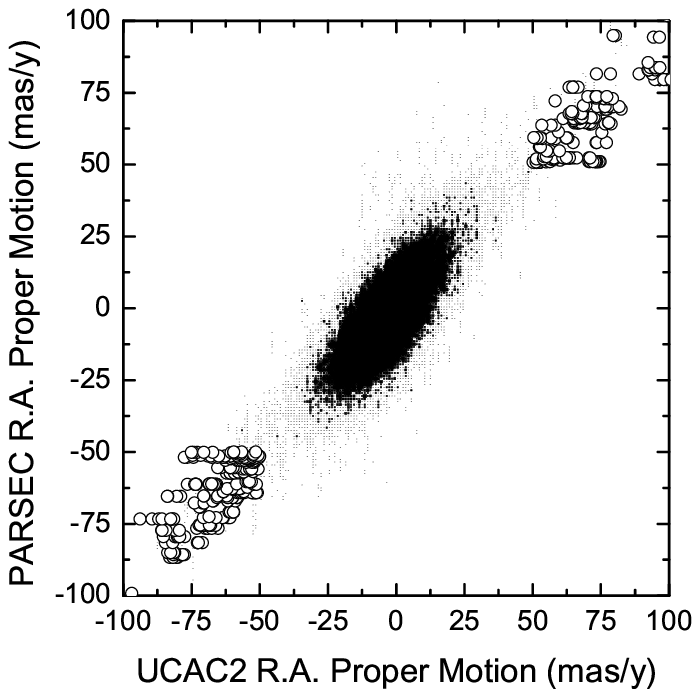}
\plotone{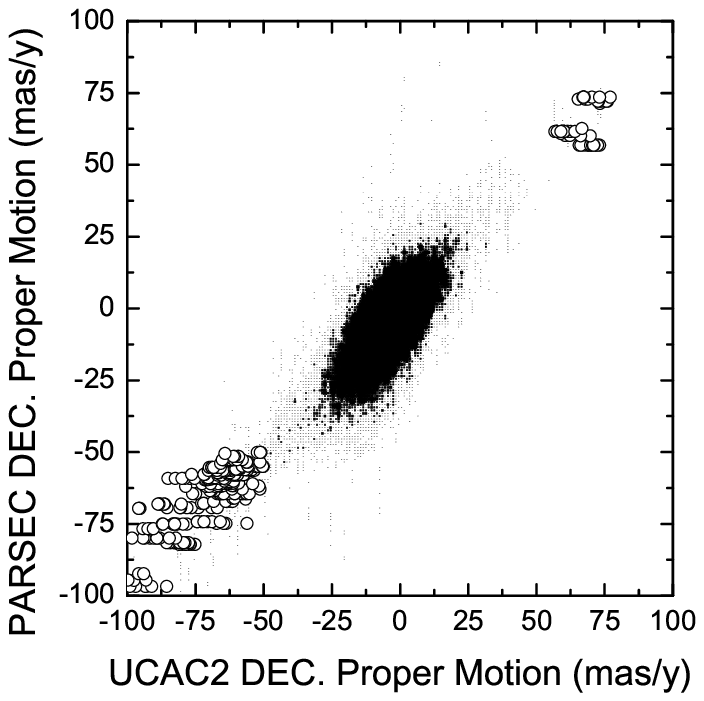}
\caption{Comparison to UCAC2 proper motions in Right Ascension top panel and
    in Declination lower panel. For the smaller proper motions, the pairs
in bins of 1mas/y are represented by dots with sizes proportional to
the counts in the bin. The pairs in which any member is larger than 
$|$50$|$mas/y are individually represented by open circles.}
  \label{Fig6a_PMRAparsecXucac2.ps}
\end{figure}
The histograms of the right ascension and declination proper motions
distributions are presented in Figure \ref{Fig9a_PMRAdistribution.ps}. The
mean value for $\alpha$ is -2.8mas/yr (standard deviation $\sigma$=12.1mas),
and -4.0mas/yr (standard deviation $\sigma$=12.3mas) for $\delta$. This
compares well with the corresponding values for the UCAC2 catalogue in the
same regions, which are, -2.7mas/yr (standard deviation $\sigma$=14.6mas) for
$\alpha$, and -3.6mas/yr (standard deviation $\sigma$=30.1mas) for
$\delta$. Zonal averages (3$^h$$\times$30$^\circ$) also produced similar means
for the PARSEC program and the UCAC2 catalogue stars. Figure
\ref{Fig6a_PMRAparsecXucac2.ps} compares the proper motions of stars in common
with the PARSEC program and the UCAC2 catalogue. The Pearson's linear
correlation coefficient is 0.95 both on right ascension and on
declination. The largest difference appear for the smallest proper motions
($\>$25mas/yr), the central parts in figure \ref{Fig6a_PMRAparsecXucac2.ps},
where the PARSEC values typically exceed those from the UCAC2 by 6\%.

\begin{figure}
\plotone{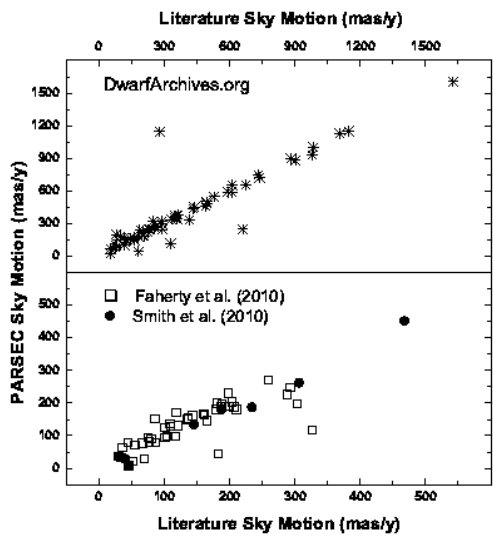} 
\caption{Comparison to literature values of the proper motions for the targets. 
         On the top panel the comparison is made for the 58 PARSEC targets found in the 
         archives of DwarfArchives.org. On the bottom panel the comparison is made for
         the most recent papers -  45 PARSEC targets in \citep{2010AJ139.176F}
         and 8 PARSEC targets in \citep{2010AJ139.1808S}. Notice the different scale range
         in the two panels}
  \label{Fig8_SkyMotionparsec-literature.ps}
\end{figure}

 Figure
\ref{Fig8_SkyMotionparsec-literature.ps} is a similar comparison of the
program target proper motions against values found in the literature. The
linear correlation is 0.82.  
There are 5 outliers: 0523-14, 0559-14, 0624-45, 1828-48, and 1956-17. 
For each of the outliers a local comparison against the UCAC2 proper motions within
the target's CCD have shown the same level of agreement found for the program
as a whole. In the case of 0523-14 and 0624-45 the original proper motion paper
\citep{2007AJ....133.2258S} does not have any special discussion of these sources. 
Our results come respectively from 2 and 3 PARSEC observations, and more data must 
be added on to reach a clearer understanding. 
 For 0559-14, the original proper motion paper \citep{2002AJ....124.1170D}
indicates that the proper motion was taken within just 2.1 years, and the results should be taken as preliminary.
For 1828-48 the original proper motion paper \citep{2004AJ....127.2856B} indicates 
troublesome observations at high air mass and under patchy skies. For 1956-17 
the original proper motion paper is from the SuperCOSMOS Sky Survey \citep{2001MNRAS.326.1279H} 
which for individual objects may have large errors while our results include 7 consistent 
PARSEC observations.

\begin{figure}
\plotone{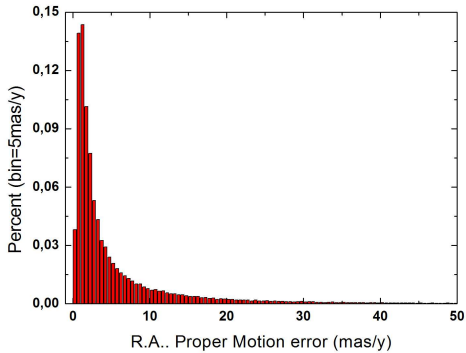}
\plotone{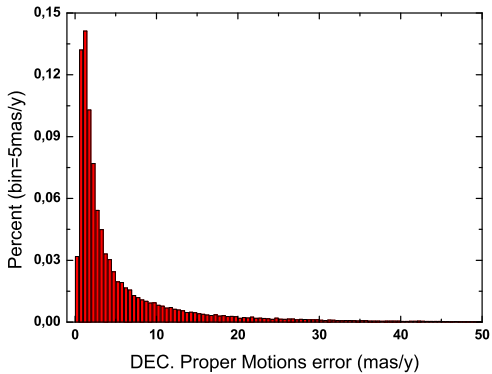}
\caption{The distribution of the right ascension and declination proper
    motion errors.}
  \label{Fig10a_PMRAerrors.ps}
\end{figure}

\begin{figure}
\plotone{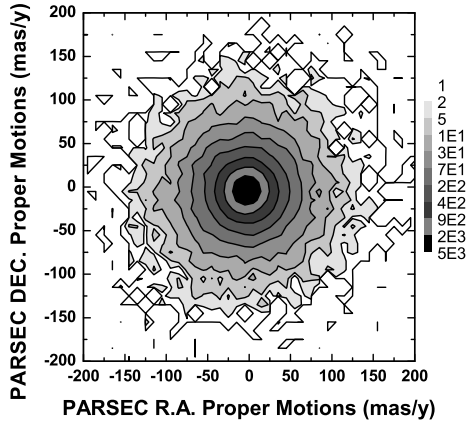}
\caption{The right ascension vs declination proper motion contour plot for 
all the stars in the PARSEC fields. }
  \label{Fig11_PMRAxPMDe.ps}
\end{figure}

The histograms of the proper motion errors are shown in Figure
\ref{Fig10a_PMRAerrors.ps}. 
The mean values are 5mas/yr both for right ascension and declination. The similarity 
of the behaviour in both coordinates, already seen in the comparison to UCAC2 values and 
for the errors distribution, is shown in Figure \ref{Fig11_PMRAxPMDe.ps}, where the 
pair-wise RA and DEC proper 
motions are plotted for all objects. The actual proper motions distribution reflects 
several factors, prominently the galactic rotation, and an uniform distrubution modulated
 by the inverse square of the distance. The combination of factors may
 result either in a Poisson-like or in a Gaussian-like distribution. We
 want to investigate whether the peculiar geometry of each CCD, and/or
 some artifact left by the astrometric reduction made indepently for each of them would
 reflect on the proper motions distribution. In order to not assume an a priori 
 model, the cumulative density distribution of the proper motions was
 fitted by an exponential decay, either along the RA or the DEC
 directions, and indeed when those are quadratically combined to produce
 the apparent sky motion. The 
exponential decays are characterized by the one free parameter which we call 
scale length.
It is an uncomplicated, robust estimator  
borrowed from the description of stochastic processes
in which events can occur continuously and are independently of one another. We
calculated the scale length in density steps of 10mas/yr in order to
investigate the presence of clumps, voids, or systematics within the parent
proper motions population. This was done for each CCD, as shown in Table
\ref{PMDIST}.  From one CCD to another the variation of the least-squares
adjusted scale length was always smaller than the rms of the adjustments, implying consistent astrometric precision from the different CCDs.

\begin{table}[ht]
  \caption{Scale length of the exponential decay of the separate and total
proper motion distributions. Values are presented for all objects, and then by CCD.
}
\begin{tabular}{lccc}
\hline  \\
CCD & \multicolumn{3} {c} { Scale Length} \\

      &  Total +/- $\sigma$ & $\mu_{\alpha} \pm \sigma$ &  $\mu_{\delta} \pm \sigma$ \\
      &     mas/yr          &     mas/yr                &     mas/yr                 \\ 
\tableline
 ALL   & 34.3 $\pm$ 7.6 & 35.9 $\pm$ 8.0 & 34.4 $\pm$ 7.7\\
 1     & 34.3 $\pm$ 7.6 & 35.9 $\pm$ 8.0 & 34.4 $\pm$ 7.7\\
 2     & 30.7 $\pm$ 6.6 & 30.6 $\pm$ 7.1 & 29.0 $\pm$ 6.5\\
 3     & 29.7 $\pm$ 6.3 & 29.6 $\pm$ 6.8 & 30.6 $\pm$ 6.7\\
 4     & 35.9 $\pm$ 8.0 & 34.4 $\pm$ 7.7 & 34.1 $\pm$ 7.8\\
 5     & 30.6 $\pm$ 7.1 & 29.0 $\pm$ 6.5 & 30.0 $\pm$ 6.9\\
 6     & 29.6 $\pm$ 6.8 & 30.6 $\pm$ 6.7 & 26.6 $\pm$ 6.2\\
 7     & 34.4 $\pm$ 7.7 & 34.1 $\pm$ 7.8 & 30.7 $\pm$ 7.3\\
 8     & 29.0 $\pm$ 6.5 & 30.0 $\pm$ 6.9 & 27.0 $\pm$ 6.8\\
\hline
\label{PMDIST}
\end{tabular}
\end{table}

The proper motion catalogue will be provided upon request 
(\footnote{The preliminary catalog can be retrieved from the PARSEC web site at http:$/$$/$parsec.oato.inaf.it$/$data$\_$releases.html}). 
We have chosen not
to distribute it to the data centers until the observations are finished and
we have a final product. The above evaluation and our first release is based on
the first year of the program, comprising 6 observation periods, from
April/2007 up to April/2008.

\begin{figure}
\plotone{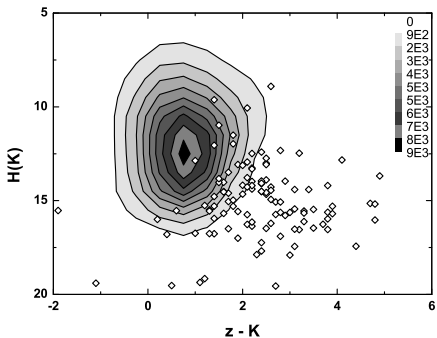}
\caption{A reduced proper motion diagram of all objects in the PARSEC proper
    motion  catalog. Diamonds are the targets and delineate the brown dwarf
    zone. The contour levels mark the quantity of field stars within the
    zone of the diagram. From the field stars furthermost in the targets
    zone possible brown dwarf candidates can be identified for spectroscopic 
    follow up.}
  \label{rpm}%
\end{figure}

In figure \ref{rpm} we plot the reduced proper motion, $H(K) = K + 5\times
log(\mu_{tot}) + 5 $, as a function of the $z-K$ color. The $z$ magnitudes
come from a zero point correction to the instrumental magnitudes of the first
observations, the $K$ are 2MASS magnitudes. The targets appear as white diamonds
and they delineate the brown dwarf zone. The field stars are represented by
contour levels, from a 100 $\times$ 100 matrix of stars count. 
From the field stars furthermost in the targets zone possible brown dwarf candidates 
can be identified for spectroscopic follow up.

\section{Conclusion}
We have presented the first parallaxes from the PARSEC program. The results
bode well for the whole program which is expected to finish in early
2011. We have confirmed a candidate binary (1004-33) and discovered another (0719-50), both of which
will make good benchmark systems. The WFI has a large field of view so we are 
able to put very good constraints on the wide binary systems and the
parallaxes allow us to immediately isolate unresolved binaries because of
their over luminosity with respect to their color. Given these two properties,
and the large sample, we expect to be able to put sensible constraints on the
binary fraction of brown dwarfs, a quantity that is critical for estimating
the substellar mass function. We have produced a catalog of proper motions
sampling the whole of the southern hemisphere. This catalogue provides an independent
validation of the UCAC2 proper motion system. The proper motion distributions
are shown to be statistically well behaved, it follows that the proper motions 
for the fainter objects will have the same precision. We will continue to update
online this catalog until the end of the program and we plan to improve it
including also GSC23 database observations.

\section{Acknowledgments}
The authors would like to acknowledge the support of: the Royal Society
International Joint Project 2007/R3; the PARSEC International Incoming
Fellowship and IPERCOOL International Research Staff Exchange Scheme within
the Marie Curie 7th European Community Framework Programme. AHA thanks CNPq
grant PQ-307126/2006-0. JIBC acknowledges CNPq financial support $\#$477943/2007-1. DNSN thanks FAPERJ grant E-26/110.177/2009.

This research has made use of: the SIMBAD database operated at CDS France; the
Second Guide Star Catalog developed as a collaboration between the Space
Telescope Science Institute and the Osservatorio Astronomico di Torino; the
Two Micron All Sky Survey which is a joint project of the University of
Massachusetts and the Infrared Processing and Analysis Center/California
Institute of Technology; and, the M, L, and T dwarf compendium housed at
dwarfArchives.org and maintained by Chris Gelino, Davy Kirkpatrick, and Adam
Burgasser.

\bibliographystyle{apj}                       
\bibliography{local.bib}

\setcounter{table}{1}
\clearpage
\begin{deluxetable}{llllllll}
\tablewidth{400pt}
\tablecaption{\label{targets} PARSEC Targets as of 1/2009}
\tablehead{
\colhead{2MASS ID}  &         
\colhead{Red. ID}      &     
\colhead{z }     &
\colhead{J }     &
\colhead{H }     &
\colhead{K$_s$ }     &
\colhead{ST}   &        
\colhead{Discovery ID} }           
\startdata
00043484-4044058 & 0004-40 & 15.8 &  13.109 & 12.055 & 11.396 &  L4.5 &       GJ 1001B, LHS 102B    \\
00062050-1720506 & 0006-17 & 18.4 &  15.662 & 14.646 & 14.010 &  L2.5 &     2MASSI J0006205-172051  \\
00100009-2031122 & 0010-20 & 16.5 &  14.134 & 13.368 & 12.882 &  L0.0 &     2MASS J00100009-2031122 \\
00135779-2235200 & 0013-22 & 18.6 &  15.775 & 14.595 & 14.036 &  L4.0 &     2MASSI J0013578-223520  \\
00145575-4844171 & 0014-48 & 16.8 &  14.050 & 13.107 & 12.723 &  L2.5 &     2MASS J00145575-4844171 \\
00165953-4056541 & 0016-40 & 18.0 &  15.316 & 14.206 & 13.432 &  L3.5 &     2MASS J00165953-4056541 \\
00300625-3739483 & 0030-37 & 17.9 &  15.204 & 14.426 & 13.885 &  L3.0 &     DENIS-P J003006.2-373948\\
00325584-4405058 & 0032-44 & 17.1 &  14.776 & 13.857 & 13.269 &  L0.0 &       EROS-MP J0032-4405    \\
00324308-2237272 & 0032-22 & 17.9 &  15.388 & 14.512 & 13.962 &  L1.0 &     2MASSI J0032431-223727  \\
00332386-1521309 & 0033-15 & 18.0 &  15.286 & 14.208 & 13.410 &  L4.0 &     2MASS J00332386-1521309 \\
00345684-0706013 & 0034-07 & 18.2 &  15.531 & 14.566 & 13.942 &  L3.0 &     2MASSI J0034568-070601  \\
00511078-1544169 & 0051-15 & 18.0 &  15.277 & 14.164 & 13.466 &  L3.5 &     2MASSW J0051107-154417  \\
00531899-3631102 & 0053-36 & 17.2 &  14.445 & 13.480 & 12.937 &  L3.5 &     2MASS J00531899-3631102 \\
00540655-0031018 & 0054-00 & 18.3 &  15.731 & 14.891 & 14.380 &  L1.0 &    SDSSp J005406.55-003101.8\\
00584253-0651239 & 0058-06 & 17.1 &  14.311 & 13.444 & 12.904 &  L0.0 &      SIPS0058-0651          \\
01090150-5100494 & 0109-51 & 14.6 &  12.228 & 11.538 & 11.092 &  L1.0 &      SIPS0109-5100          \\
01174748-3403258 & 0117-34 & 17.9 &  15.178 & 14.209 & 13.489 &  L2.0 &     2MASSI J0117474-340325  \\
01230050-3610306 & 0123-36 & 16.4 &  13.639 & 13.108 & 12.191 &  L2.0 &      2MASSJ01230050-3610306 \\
01253689-3435049 & 0125-34 & 18.3 &  15.522 & 14.474 & 13.898 &  L2.0 &     2MASSI J0125369-343505  \\
01282664-5545343 & 0128-55 & 16.6 &  13.775 & 12.916 & 12.336 &  L2.0 &      SIPS0128-5545          \\
01443536-0716142 & 0144-07 & 16.9 &  14.191 & 13.008 & 12.268 &  L5.0 &     2MASS J01443536-0716142 \\
01473282-4954478 & 0147-49 & 15.8 &  13.058 & 12.366 & 11.916 &  L2.0$^a$  &   ...   \\
02052940-1159296 & 0205-11 & 17.4 &  14.587 & 13.568 & 12.998 &  L5.5 &      DENIS-P J0205.4-1159   \\
02182913-3133230 & 0218-31 & 17.4 &  14.728 & 13.808 & 13.154 &  L3.0 &     2MASSI J0218291-313322  \\
02192807-1938416 & 0219-19 & 16.9 &  14.110 & 13.339 & 12.910 &  L2.5 &        SSSPM J0219-1939     \\
02271036-1624479 & 0227-16 & 16.1 &  13.573 & 12.630 & 12.143 &  L1.0 &     2MASS J02271036-1624479 \\
02304498-0953050 & 0230-09 & 17.7 &  14.818 & 13.912 & 13.403 &  T0.0$^a$  &   ...  \\
02355993-2331205 & 0235-23 & 15.2 &  12.690 & 12.725 & 12.186 &  L1.0 &            GJ 1048B         \\
02354756-0849198 & 0235-08 & 18.3 &  15.571 & 14.812 & 14.191 &  L2.0 &    SDSS J023547.56-084919.8 \\
02394245-1735471 & 0239-17 & 16.6 &  14.291 & 13.525 & 13.039 &  L0.0 &      SIPS0239-1735          \\
02431371-2453298 & 0243-24 & 18.9 &  15.381 & 15.137 & 15.216 &  T6.0 &     2MASSI J0243137-245329  \\
02550357-4700509 & 0255-47 & 16.1 &  13.246 & 12.204 & 11.558 &  L9.0 &       DENIS-P J0255-4700    \\
02572581-3105523 & 0257-31 & 17.6 &  14.672 & 13.518 & 12.876 &  L8.0 &     2MASS J02572581-3105523 \\
03101401-2756452 & 0310-27 & 18.5 &  15.795 & 14.662 & 13.959 &  L5.0 &     2MASS J03101401-2756452 \\
03185403-3421292 & 0318-34 & 18.5 &  15.569 & 14.346 & 13.507 &  L7.0 &     2MASS J03185403-3421292 \\
03480772-6022270 & 0348-60 & 18.8 &  15.318 & 15.559 & 15.602 &  T7.0 &     2MASS J03480772-6022270 \\
03504861-0518126 & 0350-05 & 18.8 &  16.327 & 15.525 & 15.125 &  L1.0 &    SDSS J035048.62-051812.8 \\
03572695-4417305 & 0357-44 & 16.7 &  14.367 & 13.531 & 12.907 &  L0.0 &    DENIS-P J035726.9-441730 \\
03572110-0641260 & 0357-06 & 18.3 &  15.953 & 15.060 & 14.599 &  L0.0 &    SDSS J035721.11-064126.0 \\
04082905-1450334 & 0408-14 & 16.9 &  14.222 & 13.337 & 12.817 &  L4.5 &     2MASSI J0408290-145033  \\
04234858-0414035 & 0423-04 & 17.3 &  14.465 & 13.463 & 12.929 &  L0.0 &    SDSSp J042348.57-041403.5\\
04390101-2353083 & 0439-23 & 17.3 &  14.408 & 13.409 & 12.816 &  L6.5 &     2MASSI J0439010-235308  \\
04430581-3202090 & 0443-32 & 18.0 &  15.273 & 14.350 & 13.877 &  L5.0 &     2MASSI J0443058-320209  \\
05185995-2828372 & 0518-28 & 18.8 &  15.978 & 14.830 & 14.162 &  L1.0 &     2MASS J05185995-2828372 \\
05233822-1403022 & 0523-14 & 15.9 &  13.084 & 12.220 & 11.638 &  L5.0 &     2MASSI J0523382-140302  \\
05395200-0059019 & 0539-00 & 16.7 &  14.033 & 13.104 & 12.527 &  L3.0 &      SIPS0539-0059          \\
05591914-1404488 & 0559-14 & 17.3 &  13.802 & 13.679 & 13.577 &  T4.5 &     2MASS J05591914-1404488 \\
06141196-2019181 & 0614-20 & 17.6 &  14.783 & 13.901 & 13.375 &  L4.0 &      SIPS0614-2019          \\
06244595-4521548 & 0624-45 & 17.2 &  14.480 & 13.335 & 12.595 &  L5.0 &     2MASS J06244595-4521548 \\
06395596-7418446 & 0639-74 & 18.5 &  15.795 & 14.723 & 14.038 &  L5.0 &     2MASS J06395596-7418446 \\
06411840-4322329 & 0641-43 & 16.3 &  13.751 & 12.941 & 12.451 &  L1.5 &     2MASS J06411840-4322329 \\
07193188-5051410 & 0719-50 & 16.5 &  14.094 & 13.282 & 12.773 &  L0.0 &     2MASS J07193188-5051410 \\
07291084-7843358 & 0729-78 & 18.3 &  15.440 & 14.947 & 14.635 &  L0.0$^b$  &         2\_3367 	  \\
08283419-1309198 & 0828-13 & 15.6 &  12.803 & 11.851 & 11.297 &  L2.0 &        SSSPM J0829-1309     \\
08320451-0128360 & 0832-01 & 16.6 &  14.128 & 13.318 & 12.712 &  L1.5 &     2MASSW J0832045-012835  \\
08354256-0819237 & 0835-08 & 15.9 &  13.169 & 11.938 & 11.136 &  L5.0 &     2MASSI J0835425-081923  \\
08592547-1949268 & 0859-19 & 18.4 &  15.527 & 14.436 & 13.751 &  L6.0 &     2MASSI J0859254-194926  \\
09095749-0658186 & 0909-06 & 16.2 &  13.890 & 13.090 & 12.539 &  L0.0 &       DENIS-P J0909-0658    \\
09211410-2104446 & 0921-21 & 15.5 &  12.779 & 12.152 & 11.690 &  L4.5 &     2MASS J09211410-2104446 \\
09221952-8010399 & 0922-80 & 18.1 &  15.276 & 14.285 & 13.681 &  L2.0 &     2MASS J09221952-8010399 \\
09283972-1603128 & 0928-16 & 18.1 &  15.322 & 14.292 & 13.615 &  L2.0 &     2MASSW J0928397-160312  \\
09532126-1014205 & 0953-10 & 15.8 &  13.469 & 12.644 & 12.142 &  L0.0 &     2MASS J09532126-1014205 \\
10044030-1318186 & 1004-13 & 17.6 &  14.685 & 13.883 & 13.357 &  T0.0$^a$  &   ...  \\
10043929-3335189 & 1004-33 & 17.3 &  14.480 & 13.490 & 12.924 &  L4.0 &     2MASSW J1004392-333518  \\
10185879-2909535 & 1018-29 & 16.7 &  14.213 & 13.418 & 12.796 &  L1.0 &     2MASSW J1018588-290953  \\
10452400-0149576 & 1045-01 & 15.7 &  13.160 & 12.352 & 11.780 &  L1.0 &     2MASSI J1045240-014957  \\
10473109-1815574 & 1047-18 & 17.0 &  14.199 & 13.423 & 12.891 &  L2.5 &       DENIS-P J1047-1815    \\
10584787-1548172 & 1058-15 & 16.9 &  14.155 & 13.226 & 12.532 &  L3.0 &      DENIS-P J1058.7-1548   \\
10595138-2113082 & 1059-21 & 17.1 &  14.556 & 13.754 & 13.210 &  L1.0 &     2MASSI J1059513-211308  \\
11220826-3512363 & 1122-35 & 18.1 &  15.019 & 14.358 & 14.383 &  T2.0 &     2MASS J11220826-3512363 \\
11223624-3916054 & 1122-39 & 18.4 &  15.705 & 14.682 & 13.875 &  L3.0 &     2MASSW J1122362-391605  \\
11263991-5003550 & 1126-50 & 15.9 &  13.997 & 13.284 & 12.829 &  L6.5 &     2MASS J11263991-5003550 \\
11544223-3400390 & 1154-34 & 16.6 &  14.195 & 13.331 & 12.851 &  L0.0 &     2MASS J11544223-3400390 \\
12255432-2739466 & 1225-27 & 18.8 &  15.260 & 15.098 & 15.073 &  T6.0 &     2MASS J12255432-2739466 \\
12281523-1547342 & 1228-15 & 17.2 &  14.378 & 13.347 & 12.767 &  L6.0 &      DENIS-P J1228.2-1547   \\
12462965-3139280 & 1246-31 & 18.2 &  15.024 & 14.186 & 13.974 &  T1.0$^a$  &   ...  \\
12545393-0122474 & 1254-01 & 18.0 &  14.891 & 14.090 & 13.837 &  T2.0 &    SDSSp J125453.90-012247.4\\
13262009-2729370 & 1326-27 & 18.6 &  15.847 & 14.741 & 13.852 &  L5.0 &     2MASSW J1326201-272937  \\
13314894-0116500 & 1331-01 & 18.4 &  15.459 & 14.475 & 14.073 &  L8.5 &    SDSS J133148.92-011651.4 \\
13411160-3052505 & 1341-30 & 17.3 &  14.607 & 13.725 & 13.081 &  L2.0 &     2MASS J13411160-3052505 \\
14044948-3159330 & 1404-31 & 18.8 &  15.577 & 14.955 & 14.538 &  T2.5 &     2MASS J14044941-3159329 \\
14252798-3650229 & 1425-36 & 16.5 &  13.747 & 12.575 & 11.805 &  L5.0 &   DENIS-P J142527.97-365023.\\
14385498-1309103 & 1438-13 & 18.2 &  15.490 & 14.504 & 13.863 &  L3.0 &     2MASSW J1438549-130910  \\
14413716-0945590 & 1441-09 & 16.4 &  14.020 & 13.190 & 12.661 &  L0.5 &  DENIS-P J1441-0945, G 124-6\\
14571496-2121477 & 1457-21 & 18.8 &  15.324 & 15.268 & 15.242 &  T7.5 &           Gliese 570D       \\
15074769-1627386 & 1507-16 & 15.6 &  12.830 & 11.895 & 11.312 &  L5.5 &     2MASSW J1507476-162738  \\
15200224-4422419 & 1520-44 & 16.0 &  13.228 & 12.364 & 11.894 &  L4.5 &    2MASS J15200224-4422419A \\
15230657-2347526 & 1523-23 & 17.0 &  14.203 & 13.420 & 12.903 &  L2.5 &     2MASS J15230657-2347526 \\
15302867-8145375 & 1530-81 & 17.0 &  14.154 & 13.601 & 13.404 &  L0.0$^b$  &         2\_105  	  \\
15344984-2952274 & 1534-29 & 18.4 &  14.900 & 14.866 & 14.843 &  T5.5 &     2MASSI J1534498-295227  \\
15394189-0520428 & 1539-05 & 16.6 &  13.922 & 13.060 & 12.575 &  L2.0 &   DENIS-P J153941.96-052042.\\
15474719-2423493 & 1547-24 & 16.3 &  13.970 & 13.271 & 12.742 &  L0.0 &     DENIS-P J154747.2-242349\\
15485834-1636018 & 1548-16 & 16.7 &  13.891 & 13.104 & 12.635 &  L2.0 &     2MASS J15485834-1636018 \\
16184503-1321297 & 1618-13 & 16.6 &  14.247 & 13.402 & 12.920 &  L0.0 &     2MASS J16184503-1321297 \\
16202614-0416315 & 1620-04 & 18.0 &  15.283 & 14.348 & 13.598 &  L2.5 &            GJ 618.1B        \\
16335933-0640552 & 1633-06 & 19.0 &  16.138 & 15.165 & 14.544 &  L6.0 &    SDSS J163359.23-064056.5 \\
16360078-0034525 & 1636-00 & 17.0 &  14.590 & 13.904 & 13.415 &  L0.0 &    SDSSp J163600.79-003452.6\\
16452211-1319516 & 1645-13 & 15.0 &  12.451 & 11.685 & 11.145 &  L1.5 &     2MASSW J1645221-131951  \\
17054834-0516462 & 1705-05 & 16.1 &  13.309 & 12.552 & 12.032 &  L4.0 &   DENIS-P J170548.38-051645.\\
17072343-0558249 & 1707-05 & 16.7 &  12.052 & 11.260 & 10.711 &  L3.0 &    2MASS J17072343-0558249B \\
17374334-1057425 & 1737-10 & 19.0 &  15.842 & 15.348 & 15.054 &  T2.0$^a$  &   ...  \\
17502484-0016151 & 1750-00 & 16.0 &  13.294 & 12.411 & 11.849 &  L5.5 &     2MASS J17502484-0016151 \\
17534518-6559559 & 1753-65 & 16.9 &  14.095 & 13.108 & 12.424 &  L4.0 &     2MASS J17534518-6559559 \\
18244550-7128196 & 1824-71 & 18.5 &  15.677 & 15.290 & 14.849 &  L0.0$^b$  &         2\_5716 	  \\
18283572-4849046 & 1828-48 & 18.7 &  15.175 & 14.908 & 15.181 &  T5.5 &     2MASS J18283572-4849046 \\
18401904-5631138 & 1840-56 & 18.9 &  16.066 & 15.523 & 15.186 &  L9.0$^b$ &         2\_5580 	  \\
19285196-4356256 & 1928-43 & 17.9 &  15.199 & 14.127 & 13.457 &  L4.0 &     2MASS J19285196-4356256 \\
19360187-5502322 & 1936-55 & 17.2 &  14.486 & 13.628 & 13.046 &  L5.0 &     2MASS J19360187-5502322 \\
19561542-1754252 & 1956-17 & 16.1 &  13.754 & 13.108 & 12.651 &  L0.0 &     2MASS J19561542-1754252 \\
20025073-0521524 & 2002-05 & 18.2 &  15.316 & 14.278 & 13.417 &  L6.0 &     2MASS J20025073-0521524 \\
20115649-6201127 & 2011-62 & 18.8 &  15.566 & 15.099 & 14.572 &  T1.0$^a$  &   ...  \\
20232858-5946519 & 2023-59 & 18.7 &  15.530 & 14.965 & 14.485 &  T1.0$^a$  &   ...  \\
20261584-2943124 & 2026-29 & 17.3 &  14.802 & 13.946 & 13.360 &  L1.0 &     2MASS J20261584-2943124 \\
20414283-3506442 & 2041-35 & 17.6 &  14.887 & 13.987 & 13.401 &  L2.0 &     2MASS J20414283-3506442 \\
20450238-6332066 & 2045-63 & 15.4 &  12.619 & 11.807 & 11.207 &  L4.0 &      SIPS2045-6332          \\
20575409-0252302 & 2057-02 & 15.6 &  13.121 & 12.268 & 11.724 &  L1.5 &     2MASSI J2057540-025230  \\
21015233-2944050 & 2101-29 & 18.8 &  15.604 & 14.845 & 14.554 &  T1.0$^a$  &   ...  \\
21022212-6046181 & 2102-60 & 18.8 &  15.632 & 15.200 & 14.827 &  T2.0$^a$  &   ...  \\
21041491-1037369 & 2104-10 & 16.6 &  13.841 & 12.975 & 12.369 &  L2.5 &     2MASSI J2104149-103736  \\
21075409-4544064 & 2107-45 & 17.3 &  14.915 & 13.953 & 13.380 &  L0.0 &     2MASS J21075409-4544064 \\
21304464-0845205 & 2130-08 & 16.7 &  14.137 & 13.334 & 12.815 &  L1.5 &     2MASSW J2130446-084520  \\
21324898-1452544 & 2132-14 & 19.0 &  15.714 & 15.382 & 15.268 &  T3.0$^a$  &   ...  \\
21481326-6323265 & 2148-63 & 18.3 &  15.330 & 14.338 & 13.768 &  L8.0$^a$  &   ...  \\
21501592-7520367 & 2150-75 & 16.6 &  14.056 & 13.176 & 12.673 &  L1.0 &     2MASS J21501592-7520367 \\
21574904-5534420 & 2157-55 & 17.0 &  14.263 & 13.440 & 13.002 &  L0.0 &     2MASS J21574904-5534420 \\
21580457-1550098 & 2158-15 & 17.8 &  15.040 & 13.867 & 13.185 &  L4.0 &     2MASS J21580457-1550098 \\
22041052-5646577 & 2204-56 & 16.7 &  11.908 & 11.306 & 11.208 &  T1.0 &           eps Indi Ba       \\
22064498-4217208 & 2206-42 & 18.3 &  15.555 & 14.447 & 13.609 &  L2.0 &     2MASSW J2206450-421721  \\
22092183-2711329 & 2209-27 & 18.9 &  15.786 & 15.138 & 15.097 &  T2,0$^a$  &   ...  \\
22134491-2136079 & 2213-21 & 17.9 &  15.376 & 14.404 & 13.756 &  L0.0 &     2MASS J22134491-2136079 \\
22244381-0158521 & 2224-01 & 16.9 &  14.073 & 12.818 & 12.022 &  L3.5 &     2MASSW J2224438-015852  \\
22521073-1730134 & 2252-17 & 17.2 &  14.313 & 13.360 & 12.901 &  L7.5 &   DENIS-P J225210.73-173013.\\
22545194-2840253 & 2254-28 & 16.5 &  14.134 & 13.432 & 12.955 &  L0.5 &     2MASSI J2254519-284025  \\
22552907-0034336 & 2255-00 & 18.0 &  15.650 & 14.756 & 14.437 &  L0.0 &    SDSSp J225529.09-003433.4\\
23101846-1759090 & 2310-17 & 16.9 &  14.376 & 13.578 & 12.969 &  L1.0 &        SSSPM J2310-1759     \\
23185497-1301106 & 2318-13 & 18.8 &  15.553 & 15.237 & 15.024 &  T3,0$^a$  &   ...   \\
23302258-0347189 & 2330-03 & 17.0 &  14.475 & 13.745 & 13.121 &  L1.0 &     2MASS J23302258-0347189 \\
23440624-0733282 & 2344-07 & 17.6 &  14.802 & 13.846 & 13.232 &  L4.5 &     2MASS J23440624-0733282 \\
23462656-5928426 & 2346-59 & 17.3 &  14.515 & 13.905 & 13.500 &  L5.0 &      SIPS2346-5928          \\
23515044-2537367 & 2351-25 & 14.8 &  12.471 & 11.725 & 11.269 &  L0.0 &     SIPS J2351-2537   	  \\
\enddata                                                             
\tablenotetext{a}{These objects have been provided pre-publication from  a study
  being undertaken by D. Pinfield (Univ. of
  Hertfordshire) of the Galactic Plane. The spectral types are based on photometry.}
\tablenotetext{b}{These objects have been selected from the catalog of
  \cite{2004AA...421..763P} and photometrically classified. }
\end{deluxetable}

\end{document}